\documentclass[preprint,showpacs,preprintnumbers,amsmath,amssymb]{revtex4}

\usepackage{graphicx}% Include figure files
\usepackage{dcolumn}% Align table columns on decimal point
\usepackage{bm}% bold math

\begin{document}

\title{Enhanced Kerr nonlinearity for self-action via atomic
coherence in a four-level atomic system
%Comparison between a four-level and a three-level 
%scheme for enhancing Kerr nonlinearity via atomic coherence
}
\author{Tao Hong, Michael Wong Jack, Makoto Yamashita and Takaaki Mukai}
\address{NTT Basic Research Laboratories, NTT Corporation,\\
3-1, Morinosato-Wakamiya, Atsugi-shi, Kanagawa 243-0198, Japan}
%\date{\today}

\begin{abstract}

Enhancement of optical Kerr nonlinearity for self-action 
by electro-magnetically induced transparency in a four-level 
atomic system including dephasing between the ground states 
is studied in detail by solving the density matrix equations 
for the atomic levels. We discern three major contributions, 
from energy shifts of the ground states induced 
by the probe light, to the third-order 
susceptibility in the four-level system. In this four-level 
system with the frequency-degenerate probes,
quantum interference amongst the three contributions 
can, not only enhance the third-order susceptibility 
more effectively than in the three-level system 
with the same characteristic parameters, but 
also make the ratio between its real  
and imaginary part controllable. 
Due to dephasing between the two ground states and 
constructive quantum interference,  
the most effective enhancement generally occurs at an offset 
that is determined by the atomic transition frequency difference 
and the coupling Rabi frequency.

\end{abstract}
\maketitle

\section{Introduction}

\noindent The weak nonlinear response of even the best 
materials has been a dominant limitation in experimental 
research on quantum nonlinear optics for many years. A 
number of theoretical proposals, including the creation 
of a two-photon bound state \cite{Qiao} and few-photon 
quantum solitons \cite{Quantumsoliton}, have not yet been 
experimentally realized due to the lack of large-Kerr-nonlinear 
materials. However, recent research surrounding 
electro-magnetically induced transparency (EIT) \cite{EIT}, 
which uses atomic coherence to reduce absorption, has 
opened up a completely new route to achieving large optical 
nonlinearity \cite{EITNonlinear1,EITNonlnearExperiment,
EITKerrXPM,EITPhotonSwitch,EITNonOpticsEntangle}.
An EIT medium generally possesses two important features: 
vanishing resonant absorption and, simultaneously, a refractive 
index curve with a very steep gradient \cite{EIT}. 
These two features can significantly enhance the nonlinear 
interaction strength in multi-level atomic systems. In 
addition, the latter can also significantly reduce the 
group velocity of a probe light pulse and therefore 
greatly increase the effective interaction time of 
the pulse with the medium \cite{EITPhotonSwitch,Slowlight}. 
These features may therefore enable one to use an EIT 
medium to achieve nonlinear optical processes at very 
low light intensities, or even at energies of a few 
photons per atomic cross section \cite{EITPhotonSwitch,
EITNonOpticsEntangle}. Recently, many EIT-enhanced nonlinear 
phenomena have been observed in experiments, including 
the Kerr effect\cite{EITNonlnearExperiment,
EnhanceExperiment,Focusing,Mitsunaga}. 
Of particular interest to the present work, 
Wang et al. have measured the Kerr 
nonlinear coefficient for self-phase modulation using 
a three-level system in Rubidium vapor and demonstrated 
that the nonlinear coefficient is indeed enhanced by 
EIT \cite{MinXiao}.

In addition to the scheme involving three atomic levels, 
a four-level system is also a candidate for the enhancement 
of the Kerr nonlinearity for self-phase modulation 
\cite{ImamogluX3}. It is not clear at present which, 
the three-level system or the four-level system, 
provides the most advantages for the enhancement of Kerr 
nonlinearity for self-action (self-phase modulation, two-photon
absorption) for very weak light. Or 
alternatively, because a four-level system in some sense 
contains a three-level subsystem, how does the presence of 
the fourth level effect the enhancement of the nonlinearity? 
In addition, the four-level system considered in 
Ref. \cite{ImamogluX3} did not include the dephasing 
between the two ground states. How does the dephasing, 
which is present in all realistic systems, alter the 
effective enhancement of the nonlinearity? In order to 
answer these two questions, in this paper we analyze 
a four-level system with dephasing between the ground 
states. First, in section II we quantitatively 
compare a four-level EIT scheme for self-action 
with the corresponding three-level scheme. 
We discern that there are three major 
contributions, due to energy shifts 
of the ground states induced by the probe field, to the 
third-order susceptibility in the four-level system, 
and we find that quantum interference amongst the three 
contributions can, not only enhance the third-order 
susceptibility more effectively in the four-level system 
than in the three-level system, but also make the ratio 
between its real and imaginary part controllable. 
Next, in section III we analyze the detailed behavior 
of the most effectively enhanced nonlinearity in the 
four-level system.  As a result we find that in general, 
the most effective enhancement does not occur exactly 
at the center of the transparency window due to  
quantum interference and a finite dephasing rate. 
On the contrary, it occurs at an offset that is 
determined by the atomic transition frequency difference 
and the coupling Rabi frequency. 

\section{Theoretical Model}

\begin{figure}[h]
\includegraphics[width=12cm, height=14cm, angle=270]{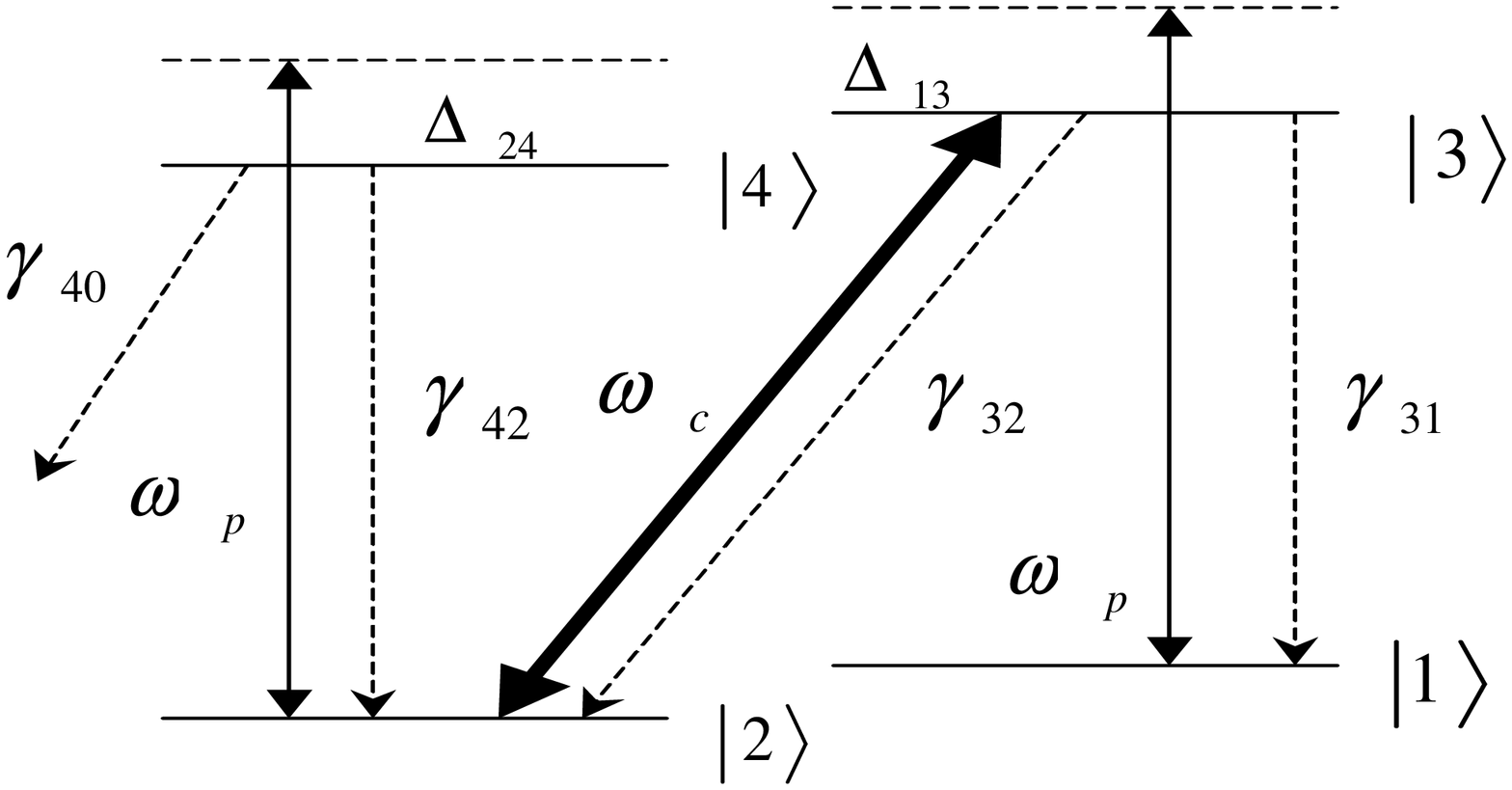}
\caption{Energy levels and optical \ couplings of the 
four-level atomic system. $\protect\omega _{c}$ is the 
angular frequency of the coupling light, and 
$\protect\omega _{p}$ the angular frequency of the 
degenerate probe light. Direct electric-dipole transition 
between two ground states, $|1\rangle $ and 
$|2\rangle $, is assumed to be forbidden. 
$\protect\gamma _{31}$, $\protect\gamma _{32}$, 
$\protect\gamma _{42}$ are decay rates from excited states 
to the ground states. $\protect\gamma _{40}$ is the decay 
rate of the state $|4\rangle $ to states other than these 
four states. $\Delta _{13}$ and $\Delta _{24}$ are
detuning frequencies of the degenerate probe light.}
\label{fig1}
\end{figure}

We consider the interaction of two light fields, a coupling 
field and a degenerate probe field, with a gas of atoms, 
as shown in Fig. \ref{fig1}. We assume the atoms can be 
described by the four-level atomic scheme. 
Consider the case when most atoms are in the ground 
state $|1\rangle $, by applying a strong coupling light 
between the states $|2\rangle$ and $|3\rangle$, we can 
dramatically reduce the resonant absorption for the weak 
probe light on the transitions $|1\rangle\leftrightarrow|3\rangle$
and $|2\rangle\leftrightarrow|4\rangle$ (see Fig. \ref{fig1}). 
Under the rotating wave approximation, 
this four-level system can be described by the following 
density matrix equations in a frame rotating at 
frequency $\omega_{p}$\cite{QuantumOptics}: 
\begin{widetext}
\begin{eqnarray}
\stackrel{\cdot }{\rho }_{11} &=&\gamma _{31}\rho _{33}+\frac{i}{2}[\Omega
_{13}\rho _{31}-\Omega _{13}^{\ast }\rho _{13}]  \label{eq1} \\
\stackrel{\cdot }{\rho }_{21} &=&[i(\Delta _{13}-\Delta _{23})-\gamma
_{21}]\rho _{21}+\frac{i}{2}[\Omega _{23}\rho _{31}+\Omega _{24}\rho
_{41}-\Omega _{13}^{\ast }\rho _{23}]  \label{eq2} \\
\stackrel{\cdot }{\rho }_{31} &=&[i\Delta _{13}-\frac{1}{2}(\gamma
_{31}+\gamma _{32})]\rho _{31}+\frac{i}{2}[\Omega _{13}^{\ast }(\rho
_{11}-\rho _{33})+\Omega _{23}^{\ast }\rho _{21}]  \label{eq3} \\
\stackrel{\cdot }{\rho }_{41} &=&[i(\Delta _{24}+\Delta _{13}-\Delta _{23})-%
\frac{1}{2}(\gamma _{40}+\gamma _{42})]\rho _{41}+\frac{i}{2}[\Omega
_{24}^{\ast }\rho _{21}-\Omega _{13}^{\ast }\rho _{43}]  \label{eq4} \\
\stackrel{\cdot }{\rho }_{22} &=&\gamma _{32}\rho _{33}+\gamma _{42}\rho
_{44}+\frac{i}{2}[\Omega _{23}\rho _{32}+\Omega _{24}\rho _{42}-\Omega
_{23}^{\ast }\rho _{23}-\Omega _{24}^{\ast }\rho _{24}]  \label{eq5} \\
\stackrel{\cdot }{\rho }_{32} &=&[i\Delta _{23}-\frac{1}{2}(\gamma
_{31}+\gamma _{32})]\rho _{32}+\frac{i}{2}[\Omega _{13}^{\ast }\rho
_{12}+\Omega _{23}^{\ast }(\rho _{22}-\rho _{33})-\Omega _{24}^{\ast }\rho
_{34}]  \label{eq6} \\
\stackrel{\cdot }{\rho }_{42} &=&[i\Delta _{24}-\frac{1}{2}(\gamma
_{40}+\gamma _{42})]\rho _{42}+\frac{i}{2}[\Omega _{24}^{\ast }(\rho
_{22}-\rho _{44})-\Omega _{23}^{\ast }\rho _{43}]  \label{eq7} \\
\stackrel{\cdot }{\rho }_{33} &=&-(\gamma _{31}+\gamma _{32})\rho _{33}+%
\frac{i}{2}[\Omega _{13}^{\ast }\rho _{13}+\Omega _{23}^{\ast }\rho
_{23}-\Omega _{13}\rho _{31}-\Omega _{23}\rho _{32}]  \label{eq8} \\
\stackrel{\cdot }{\rho }_{43} &=&[i(\Delta _{24}-\Delta _{23})-\frac{1}{2}%
(\gamma _{31}+\gamma _{32}+\gamma _{40}+\gamma _{42})]\rho _{43}+\frac{i}{2}[%
\Omega _{24}^{\ast }\rho _{23}-\Omega _{13}\rho _{41}-\Omega _{23}\rho _{42}]
\label{eq9} \\
\stackrel{\cdot }{\rho }_{44} &=&-(\gamma _{40}+\gamma _{42})\rho _{44}+%
\frac{i}{2}[\Omega _{24}^{\ast }\rho _{24}-\Omega _{24}\rho _{42}]
\label{eq10}
\end{eqnarray}
\end{widetext}
where $\rho _{ij}$ is the density matrix element, 
$\Omega _{13}=\mu_{13}E_{p0}/\hbar $ and $\Omega _{24}
=\mu _{24}E_{p0}/\hbar $ are complex Rabi frequencies for 
the probe light field amplitude $E_{p0}$, and 
$\Omega _{23}=\mu _{23}E_{c0}/\hbar $ is the complex Rabi 
frequency for the coupling light with electric field 
amplitude $E_{c0}$, where $\mu _{13}$ and $\mu _{24}$ are 
electric dipole matrix elements. $\gamma _{21}$ is the
dephasing rate between the ground states $|1\rangle $ and 
$|2\rangle $ (This was not included in Ref. \cite{ImamogluX3}). 
The detuning angular frequencies are given 
by $\Delta _{13}=\omega_{p}-\omega _{31}$, 
$\Delta _{23}=\omega _{c}-\omega _{32}$, 
$\Delta_{24}=\omega _{p}-\omega _{42}$, where $\omega _{31}$, 
$\omega _{32}$, $\omega _{42}$ are the atomic transition 
frequencies. Additionally, we assume the probe light is 
very weak, and the coupling light is on resonance 
($\Delta _{23}=0$). Under this assumption, there is a much 
larger probability of the atoms being in the ground state 
$|1\rangle $ than in other states, i.e., $\rho _{11}\approx 1$. 
Because the evolution of the atoms is very fast and the 
light is normally approximated as a continuous wave, we 
can consider the atoms to be in steady states. For 
simplicity of discussion, we assume uniform decay 
rates and uniform electric dipole matrix elements, 
i.e., $\gamma _{31}=\gamma _{32}=\gamma
_{40}=\gamma _{42}=\gamma $ and $\mu _{13}=\mu _{24}=\mu $.

Due to the degeneracy of the probe frequency, the polarization 
induced by the probe field is a superposition of the two 
off-diagonal density matrix elements,
\begin{equation}
P=n(\mu _{13}\rho _{31}+\mu _{24}\rho _{42})e^{-i\omega _{p}t}+c.c.
\end{equation}
where $n$ is the atom density. It is worth noting that this 
linear superposition gives rise to quantum interference between 
the two off-diagonal density matrix elements, because the two quantities
are complex functions and there is coherence among the atomic states.
For example, if the transition amplitudes $\rho_{31}$ and $\rho_{42}$
are in phase, an effective two-photon transition will be enhanced; 
on the other hand, if the two amplitudes are out of phase, 
then a photon emitted on one transition will be absorbed on 
the other and the effective two-photon transition will be suppressed. 
It is evident that there is no interference between the two 
off-diagonal density matrix elements in the four-level system 
with frequency-non-degenerate probes \cite{EITKerrXPM}.
Thus this superposition of the two off-diagonal density matrix 
elements is a unique feature of the four-level system 
with one frequency-degenerate probe. Because the probe field 
is monochromatic, the corresponding susceptibility has 
the simple form,
\begin{widetext} 
\begin{equation}
\varepsilon _{0}\left( \chi ^{(1)}(E_{c0})E_{p0}+\chi
^{(NL)}(E_{c0},E_{p0})E_{p0}^{3}\right) =2n(\mu _{13}\rho _{31}+\mu _{24}\rho
_{42})  \label{susceptibility}
\end{equation}
\end{widetext}
where $\chi ^{(1)}$ is the linear susceptibility and 
$\chi ^{(NL)}$ the nonlinear susceptibility. When 
$E_{p0}\rightarrow 0$, $\chi^{(NL)} $ corresponds to 
the third-order susceptibility $\chi ^{(3)}$. The
real part of $\chi ^{(3)}$ is proportional to the Kerr 
refractive index and the imaginary part of $\chi ^{(3)}$ 
to the two-photon absorption coefficient. Here, $\chi ^{(2)}$ does not 
exist because of the symmetry of the atomic medium. 
Although the occupation probabilities of atoms on both the 
states $|2\rangle$ and $|4\rangle$ are very small,
the contribution of $\rho_{42}$ to the third-order susceptibility 
is not small, as we will see in next section, therefore
neglecting the contribution of $\rho_{42}$ to the third-order
susceptibility is incorrect \cite{ImamogluX3}. 
To analyze the enhanced optical nonlinearity of the EIT medium, 
we use a numerical method, Gaussian elimination, 
to solve Eqs. (\ref{eq2})-(\ref{eq10}) in the steady state and 
then extract the first-order and the third-order susceptibilities 
via the relation (\ref{susceptibility}).

\section{Numerical comparison of four-level and three-level 
systems}

In this section, we numerically compare a four-level EIT system with a three-level EIT system. The three-level system we consider is in fact a special case of the present four-level one. When the atomic transition frequency difference $\omega _{31}-\omega _{42}$ is very large, if the probe frequency $\omega_{p}$ is close to the atomic transition frequency $\omega _{31}$, it will be far detuned from the other atomic transition frequency $\omega _{42}$. In this case, any effect from the state $|4\rangle$ is negligible, and the four-level system can be approximated as a three-level system with similar parameters. As the three-level system is actually a subsystem of the four-level system, the comparison with the three-level system is not only to show the advantage of the four-level system in enhancing Kerr nonlinearity, but also to show the contribution of the three-level subsystem, i.e., the states $|1\rangle$, $|2\rangle$ and $|3\rangle$, to the third-order susceptibility in the four-level system. 
 
We do the comparison first in terms of third-order susceptibility $\chi^{(3)}$, and then in terms of the ratio between the third-order susceptibility and the first-order susceptibility, i.e., $\chi^{(3)}/{\rm Im}\chi^{(1)}$. In the first comparison, we show that there are three major contributions of light shifts of the ground states to $\chi^{(3)}$ in the four-level system: one of them is actually produced within the three-level subsystem, and other two are specific to the four-level system. In particular, we show that there is quantum interference, which does not exist in the four-level system with frequency-non-degenerate probes \cite{EITKerrXPM}, amongst the three contributions due to atomic coherence. As a result, one can achieve not only larger third-order susceptibility in the four-level system than in the three-level system, but also a control of the ratio between its real and imaginary part. Then in the final part of the section, we consider the finite dephasing rate between the ground states, and further show the advantage of the four-level system in effectively enhancing the third-order nonlinear susceptibility with this new criterion.

\subsection{Analysis in terms of $\chi^{(3)}$}

\begin{figure}[h]
\includegraphics[width=6cm, height=8cm, angle=270]{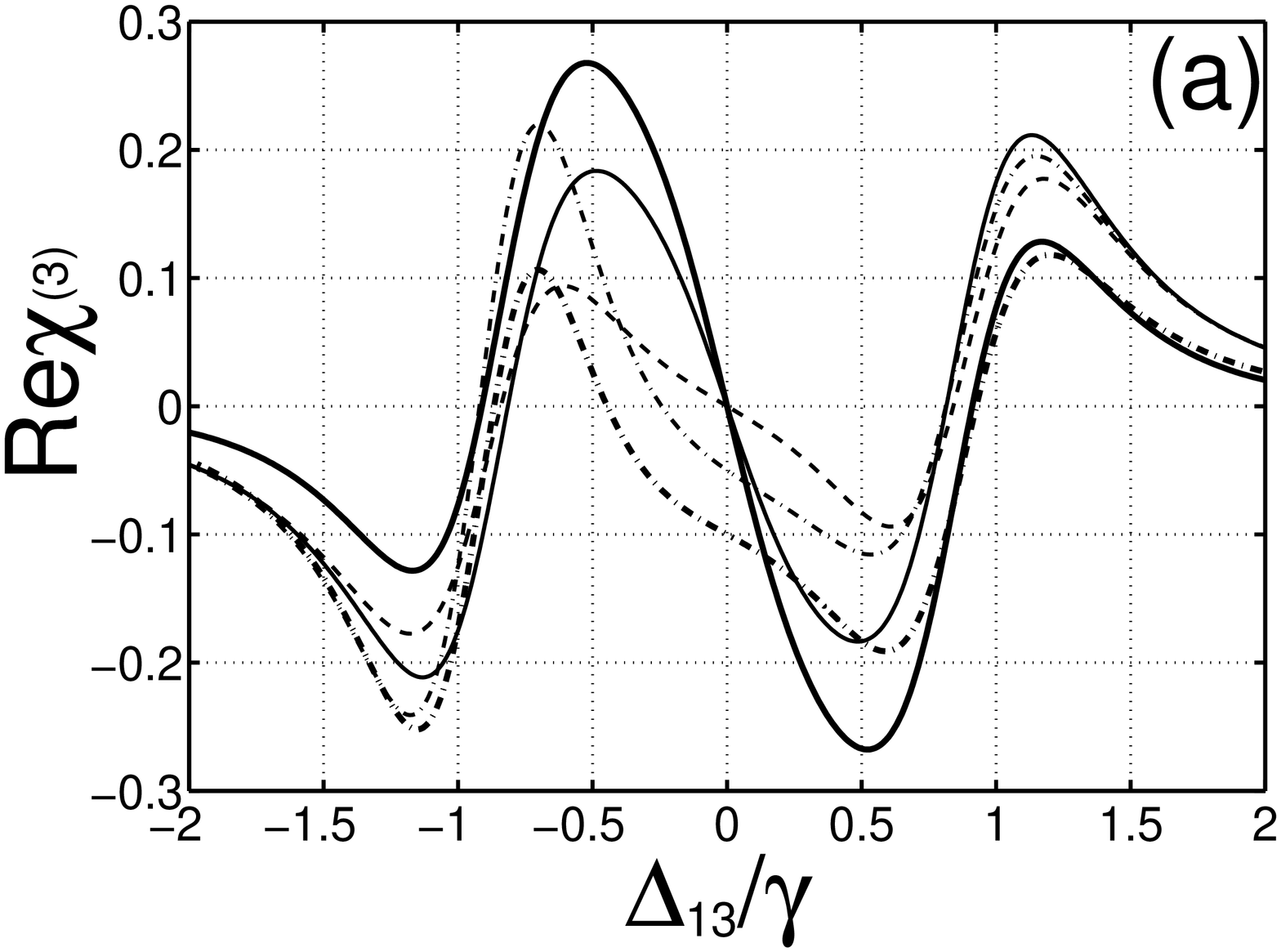}
\includegraphics[width=6cm, height=8cm, angle=270]{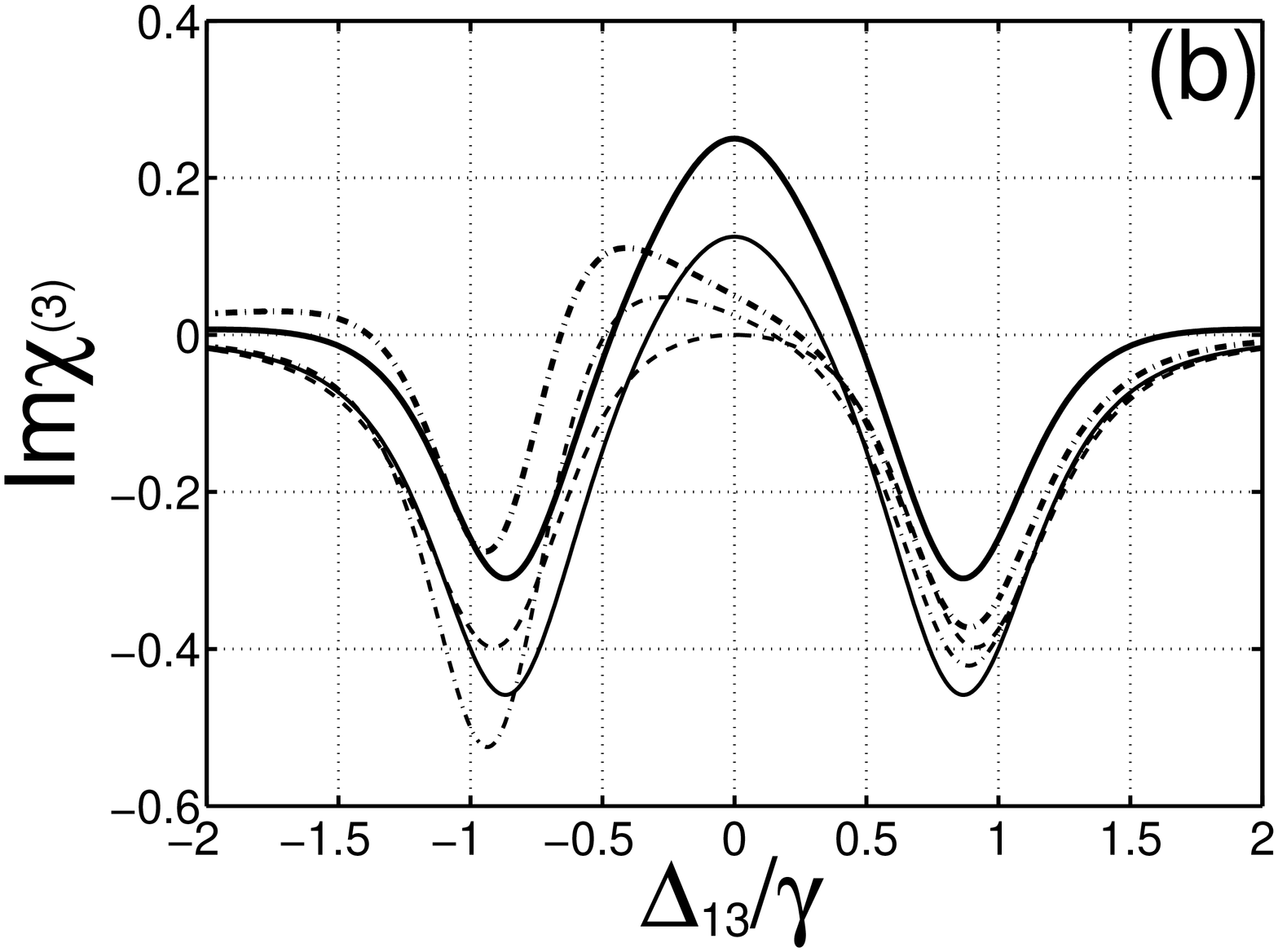}
\includegraphics[width=6cm, height=8cm, angle=270]{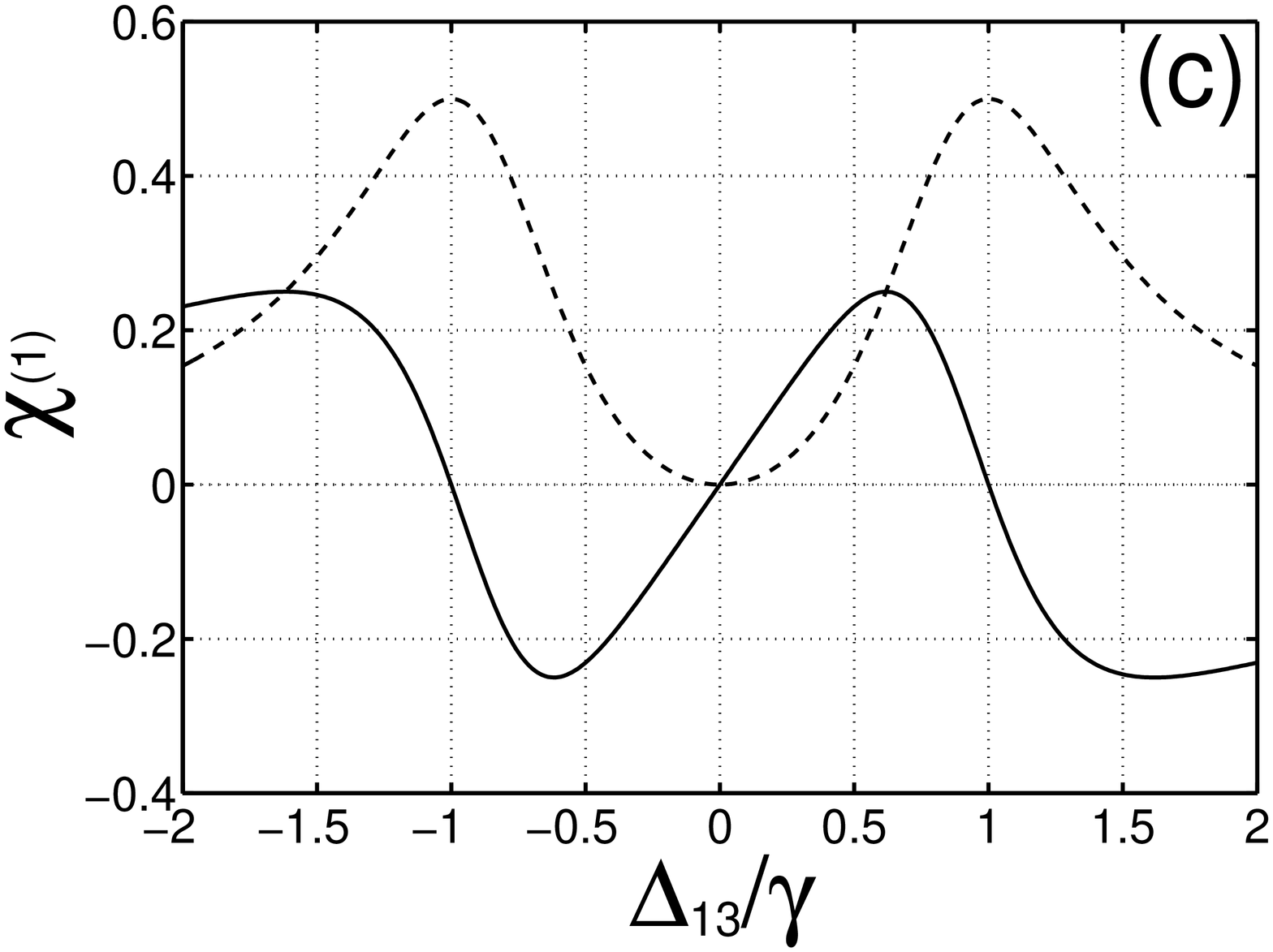}
\caption{Probe susceptibilities vs the probe detuning 
frequency. (a) and (b) show the real and imaginary parts of 
the third-order susceptibility $\chi^{(3)}$ for the
four-level system in three cases: 
$\protect\omega _{31}-\protect\omega _{42}=-10^{5}\protect\gamma $ 
(dashed lines, an approximation to a three level system), 
$\protect\omega _{31}=\protect\omega _{42}$ (thick solid lines) 
and $\protect\omega _{31}-\protect\omega _{42}=2\protect\gamma $
(thick dot-dashed lines). $\chi^{(3)}$ is in units of
 $2n\protect\mu^{4}/(\protect\varepsilon _{0}
\hbar^{3}\protect\gamma ^{3})$. 
In addition, they also show the solo contribution of $\rho_{31}$ 
to $\chi^{(3)}$ in the second case (thin solid lines) and 
the third case (thin dot-dashed lines). 
 (c) shows the transparency window: the real part (solid line) 
and imaginary part (dashed line) of the first-order 
susceptibility $\chi^{(1)}$ both become zero at $\Delta _{13}=0\gamma$. 
$\chi^{(1)}$ is in units of 
$2n\protect\mu^{2}/(\protect\varepsilon _{0}\hbar \protect\gamma )$.
In the calculation, $\protect\gamma _{21}=0\gamma$ and 
$\Omega _{23}=2\protect\gamma $. }
\label{fig2}
\end{figure}

First, following the idea described in the beginning of this section, we let $\omega_{31}-\omega_{42}=-10^{5}\gamma$ and the four-level system becomes effectively a three-level system. Any contribution of the state $|4\rangle$ to the susceptibility is negligible, so the third-order susceptibility $\chi^{(3)}$, as shown by the dashed lines in Fig. \ref{fig2} (a) and (b), only originates from the three states $|1\rangle$, $|2\rangle$ and $|3\rangle$, through the off-diagonal density matrix element $\rho_{31}$. Its real part has already been demonstrated to be much larger than that of the resonantly enhanced susceptibility without the coupling field at a finite detuning \cite{MinXiao}. 
Here we can understand the unusually large magnitude of ${\rm Re}\chi^{(3)}$ in a similar way to Ref. \cite{EITNonOpticsEntangle}: the presence of the weak probe field between the states $|1\rangle$ and $|3\rangle$ causes an energy shift of the state $|2\rangle$, which results in an effective shift of the linear susceptibility $\chi^{(1)}$ curve (see Fig. \ref{fig2} (c)). However, here there is an important difference from Ref. \cite{EITNonOpticsEntangle}: these shifts of the curves are not horizontal shifts along the frequency axis, but changes of the gradients of the curves. For ${\rm Re}\chi^{(1)}$, the gradient variation is equivalent to a small rotation of the solid curve around the zero detuning. Then we can understand that although ${\rm Re}\chi^{(3)}$ in the situation of EIT is much larger than the resonantly enhanced ${\rm Re}\chi^{(3)}$ without the coupling field at a finite detuning, the magnitudes of ${\rm Re}\chi^{(3)}$ and ${\rm Im}\chi^{(3)}$ also become zero at the zero detuning, as shown in Fig. \ref{fig2} (a) and (b). Thus the light shift within the three-level subsystem produces a contribution to the third-order susceptibility in the four-level system, and the contribution becomes important when the probe detuning is finite.
 
Next, let us consider the solo contribution of $\rho_{31}$ to $\chi^{(3)}$ at a small atomic transition frequency difference $\omega_{31}-\omega_{42}$ plotted by the thin solid lines and the thin dot-dashed lines in Fig. \ref{fig2} (a) and (b). We can find that although both the thin solid curves and the thin dot-dashed curves keep some resemblance of the dashed curves of the three-level system, the magnitudes of some parts of the curves are significantly larger than those of the three-level system at the same frequency. These variations indicate that the presence of the state $|4\rangle$ has important influence on $\chi^{(3)}$ through $\rho_{31}$. We can understand this effect in the same way as Ref. \cite{EITKerrXPM,EITNonOpticsEntangle}: the presence of the weak probe field between the states $|2\rangle$ and $|4\rangle$ leads to another energy shift of the state $|2\rangle$, which results in another effective shift of the linear susceptibility $\chi^{(1)}$, i.e., the curves experience additional shifts besides the gradient shifts produced within the three-level subsystem. For the real part of the susceptibility, the shift is mainly along the frequency axis when the atomic transition frequency difference $\omega_{31}-\omega_{42}\neq 0$. Thus the shift results in a finite value of ${\rm Re}\chi^{(3)}$ at the center of the transparency window.  

Next, following Eq. (\ref{susceptibility}), we include the contribution of $\rho_{42}$ in the calculation of $\chi^{(3)}$, as shown by the thick solid curves and thick dot-dashed curves in Fig. \ref{fig2} (a) and (b). In comparison with the corresponding thin solid curves and thin dot-dashed curves, we can find that both ${\rm Re}\chi^{(3)}$ and ${\rm Im}\chi^{(3)}$ are drastically changed once again. In particular, ${\rm Im}\chi^{(3)}$ is twice its previous value at zero detuning when $\omega_{31}=\omega_{42}$, as shown by the two solid lines in Fig. \ref{fig2} (b). This indicates that the contribution of $\rho_{42}$ to $\chi^{(3)}$ is not small, and neglecting its contribution as in Ref. \cite{ImamogluX3} does not give us correct values of $\chi^{(3)}$. We can see that the shift results in another increase of the magnitude of ${\rm Re}\chi^{(3)}$ at the center of the transparency window when the atomic transition frequency difference $\omega_{31}-\omega_{42}\neq 0$. Similarly, the energy shift of the state $|2\rangle$ due to the interaction of the weak probe field with the states $|1\rangle$ and $|3\rangle$ leads to another effective shift of $\chi^{(1)}$ through $\rho_{42}$.

\begin{figure}[h]
\includegraphics[width=9cm, height=12cm, angle=270]{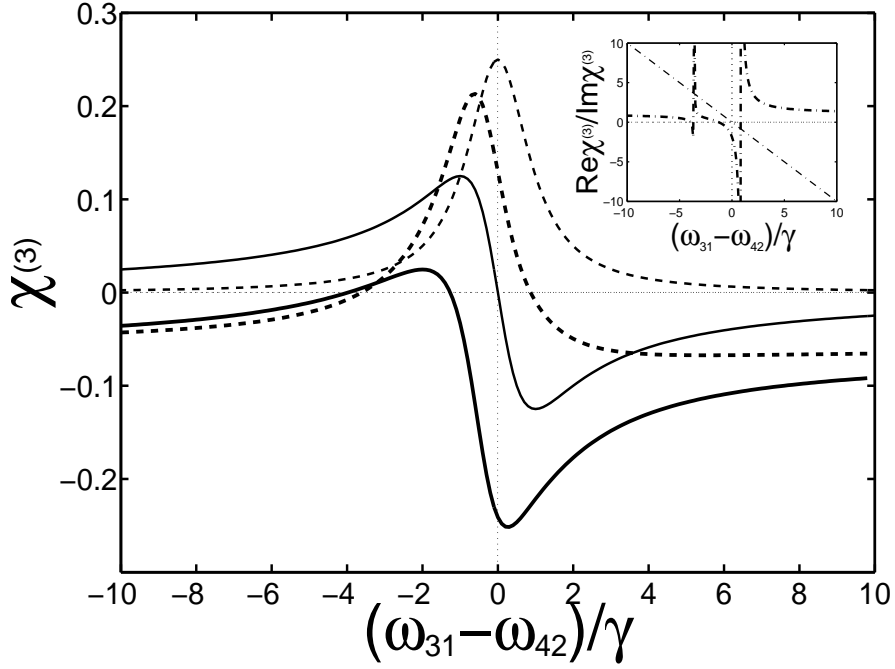}
\caption{The third-order susceptibility $\chi^{(3)}$ vs 
the atomic transition frequency difference 
$\protect\omega _{31}-\protect\omega _{42}$ in the 
four-level system for two kinds of probe detunings $\Delta_{13}$. 
The real and imaginary parts of 
$\chi^{(3)}$ in units of $2n\protect\mu ^{4}/(\protect
\varepsilon _{0}\hbar^{3}\protect\gamma ^{3})$
are represented by the solid line and the dashed line, 
respectively. The inset shows the ratio between 
the real part and the imaginary part
of $\chi^{(3)}$ as a function of 
$\omega_{31}-\omega_{42}$. In the calculation, 
$\Delta _{24}=\protect\omega_{31}-\protect\omega _{42}$, 
$\Omega _{23}=2\protect\gamma $ and $\Delta_{13}=0\protect\gamma$ 
 (thin lines) or $0.4\protect\gamma$ (thick lines). }
\label{fig3}
\end{figure}

So far we have found three major contributions of light shifts to the third-order susceptibility in the four-level EIT system. However, the most important feature is that there is quantum interference amongst the three contributions because of the existence of the atomic coherence among the atomic levels. The linear superposition of $\rho_{31}$ and $\rho_{42}$ in Eq. (\ref{susceptibility}) shows explicitly that the interference can occur. 
Through comparison of the curves of $\chi^{(3)}$ for different contributions in Fig. \ref{fig2} (a) and (b), we can also discern the variation of the susceptibility due to the interference: the magnitude of $\chi^{(3)}$ may be much increased at certain detuning due to constructive interference, but much reduced at another detuning due to destructive interference. For example, for $\Delta_{13}\approx 0.25\gamma$, the magnitude of ${\rm Re}\chi^{(3)}$, plotted by the thick dot-dashed curve in Fig. \ref{fig2} (a), is increased in comparison with the thin dot-dashed curve and the dashed curve at the same frequency due to constructive interference; however, the magnitude of ${\rm Im}\chi^{(3)}$ vanishes at the same detuning due to destructive interference, as shown in Fig. \ref{fig2} (b).  In the four-level system with frequency-non-degenerate probes, there is no such quantum interference effect \cite{EITKerrXPM}. This quantum interference process is a unique feature of the four-level system with  degenerate probes that we study in this paper. By using this quantum interference we can, not only enhance $\chi^{(3)}$ in the four-level system more than that in the three-level system, but also control the ratio between ${\rm Re}\chi^{(3)}$ and ${\rm Im}\chi^{(3)}$, as shown in Fig. \ref{fig3}.  Comparing the thin curves and the corresponding thick curves, we find that a small probe detuning can produce some zero points for ${\rm Im}\chi^{(3)}$ and change the ratio dramatically. If, for example, the four levels are magnetic sub-levels in experiments, one can use a magnetic field to control the atomic transition frequency difference $\omega_{31}-\omega_{42}$ and simultaneously set the probe detuning $\Delta_{31}$ to achieve expected ratios. Undoubtedly, such control of the nonlinear susceptibility will be very useful in a practical design of various optical devices.

\subsection{Analysis in terms of $\chi^{(3)}/{\rm Im}\chi^{(1)}$}

\begin{figure}[h]
\includegraphics[width=5cm, height=8cm, angle=270]{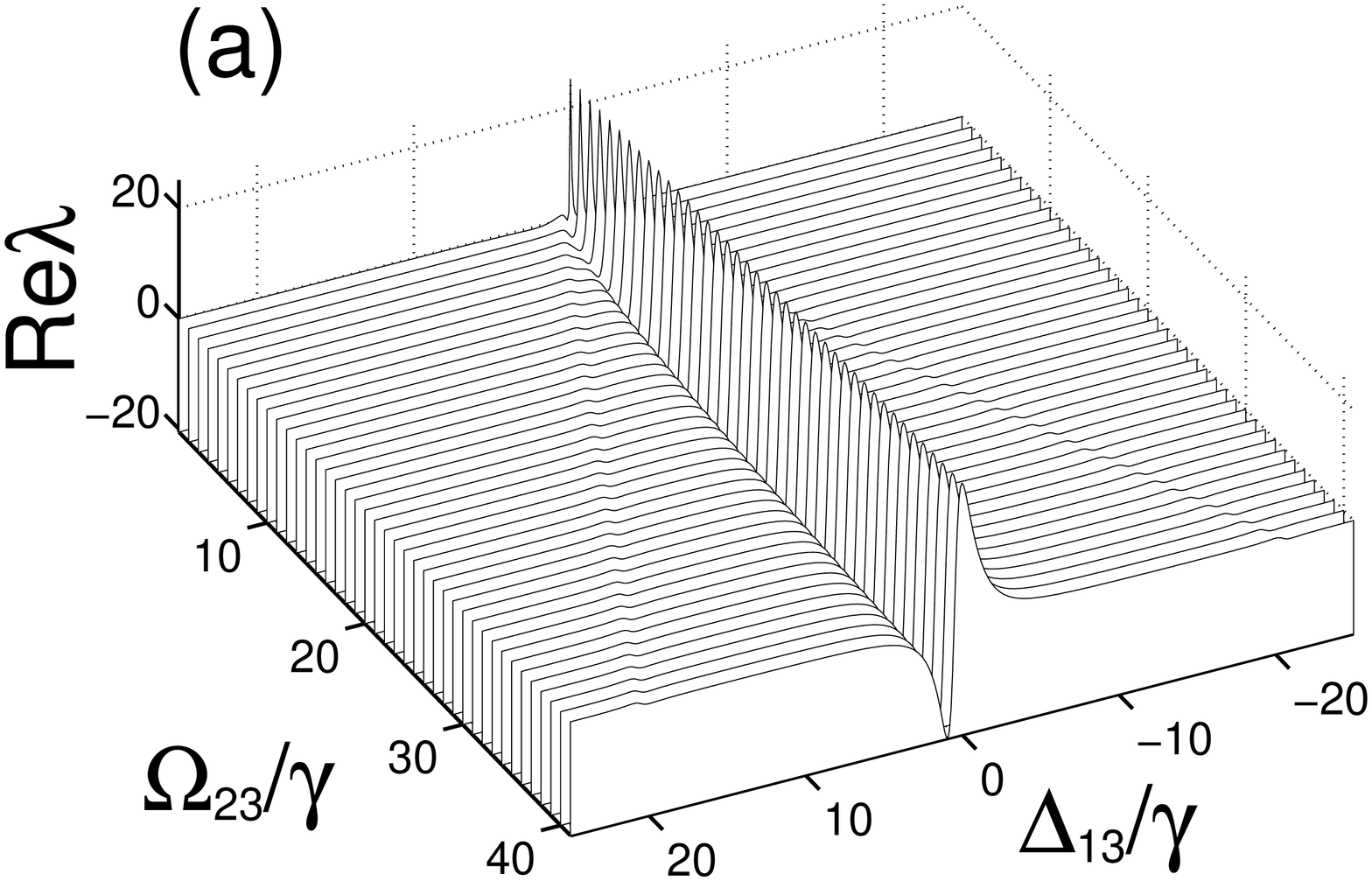}
\includegraphics[width=5cm, height=8cm, angle=270]{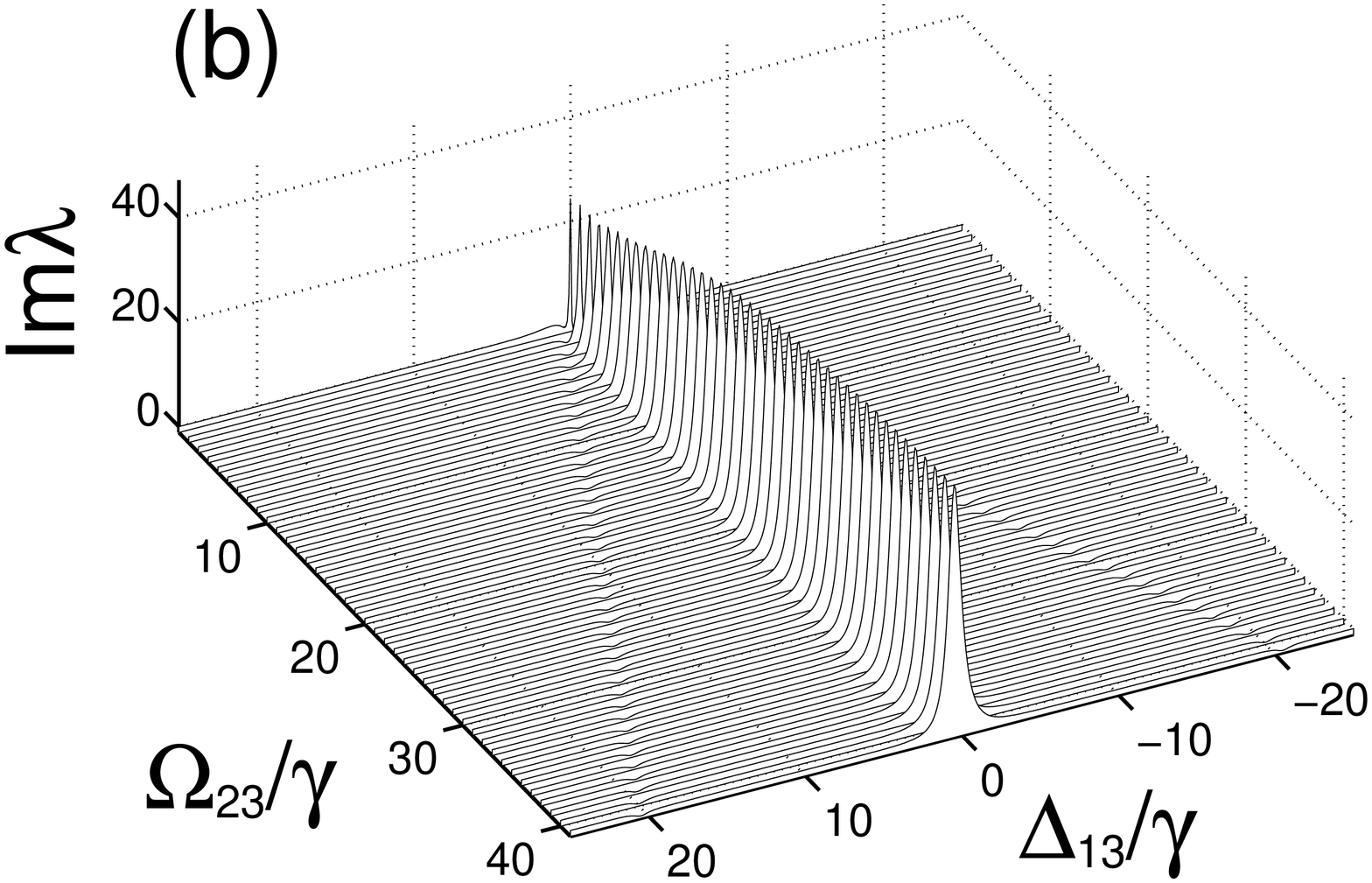}
\includegraphics[width=5cm, height=8cm, angle=270]{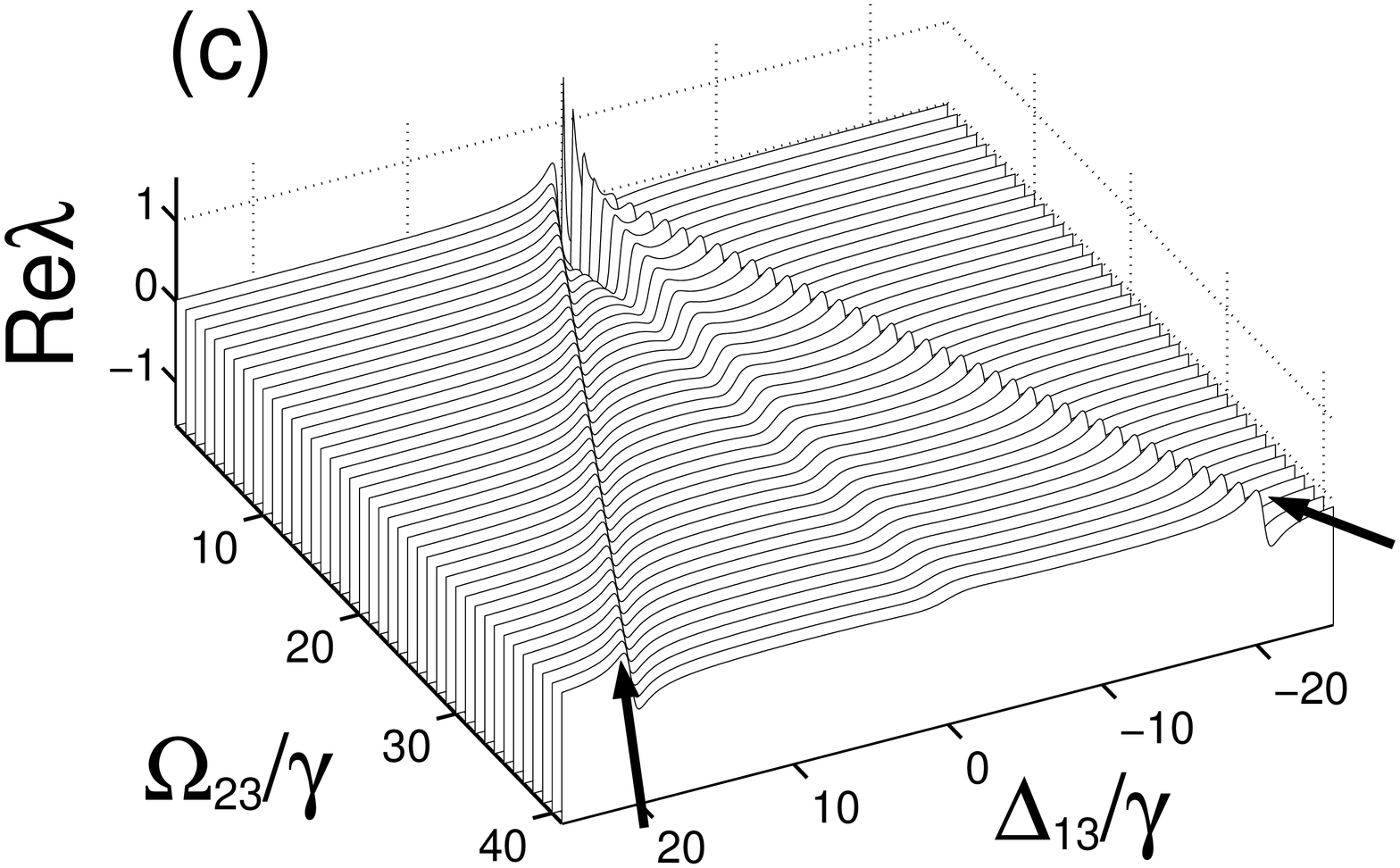}
\includegraphics[width=5cm, height=8cm, angle=270]{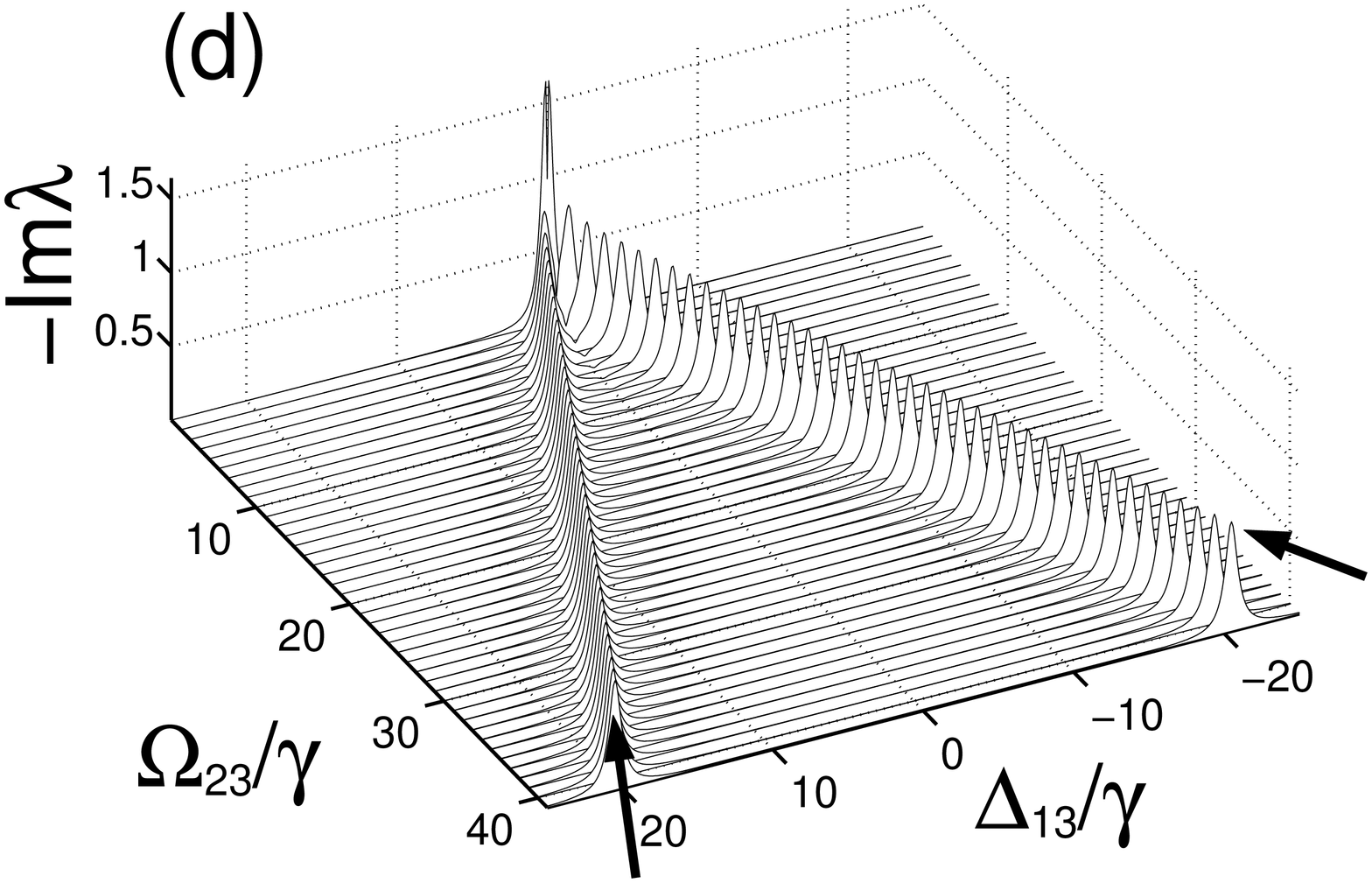}
\includegraphics[width=5cm, height=8cm, angle=270]{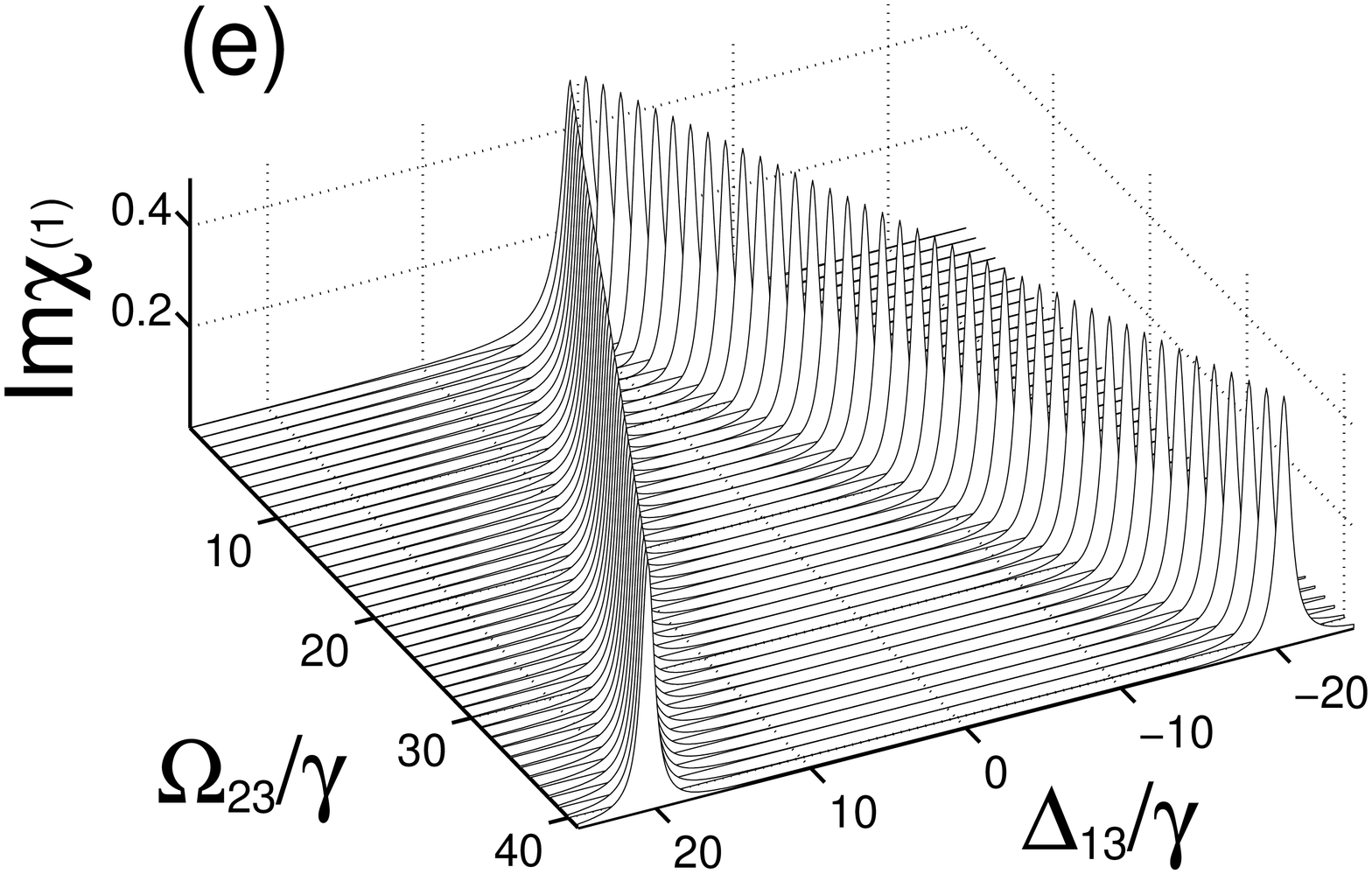}
\caption{Ratio $\lambda$ vs the probe detuning $\Delta_{13}$ 
and the Rabi frequence $\Omega_{23}$.
(a) and (b) show the ratio for the four-level system with 
$\omega _{31}-\omega _{42}=-\gamma$. 
(c) and (d) show the ratio for the three-level system, i.e.,
 the approximation of the four-level system with 
$\omega _{31}-\omega _{42}=-10^{5}\gamma$. 
 (e) shows the imaginary part of the first-order susceptibility 
$\chi^{(1)}$. The ratio $\lambda$ is in units
of $\mu^{2}/(\hbar \gamma) ^{2}$, and the linear susceptibility 
is in units of $2n\mu ^{2}/(\varepsilon _{0}\hbar \gamma )$. 
In this calculation, $\gamma _{21}=0.01\gamma $. }
\label{fig4}
\end{figure}

Because the magnitude of a nonlinear effect per unit length 
becomes very small at very low light intensities, such as when 
there are few photons per atomic  cross section, propagating light 
beams for a long distance in a nonlinear medium is usually 
a good way to magnify the nonlinear effect. However, the 
distance that light beams can propagate is limited by the 
linear absorption of the medium, which cannot be zero even under 
the conditions of EIT because normally $\gamma_{21}$ is not zero.
Thus, the ratio between the nonlinear coefficient and the linear 
absorption of the medium is the real criterion for the 
effectiveness of the enhancement of the nonlinear coefficient. 
We therefore calculate the ratio between the third-order 
susceptibility $\chi^{(3)}$ and the imaginary part of the 
first-order susceptibility $\chi^{(1)}$:
\begin{equation}
\lambda=\chi^{(3)}/{\rm Im}\chi^{(1)}
\end{equation}
The results for the four-level system and the three-level 
system are shown in Figs. \ref{fig4}(a), (b), (c), and (d). 
In order to determine the positions of the peaks of $\lambda$ 
relative to the transparency window, we also depict ${\rm Im}\chi^{(1)}$ 
in Fig. \ref{fig4}(e).

We find that for the four-level system with a small atomic
 transition frequency difference
 $\omega _{31}-\omega _{42}=-\gamma$, the position of the
 largest peak (along the detuning axis) of ${\rm Re}\lambda$ 
(or ${\rm Im}\lambda$) is always close 
to the center of the transparency window independent of the 
value of the Rabi frequency $\Omega_{23}$, as shown in 
Figs. \ref{fig4}(a) and (b). In contrast, the three-level 
system can only produce very small peaks
(along the detuning axis) of ${\rm Re}\lambda$ at the central 
part of the transparency window, as shown in
Figs. \ref{fig4}(c) and (d). The largest peaks (along the 
detuning axis) of ${\rm Re}\lambda$ (or $-{\rm Im}\lambda$) 
of the three-level system are not in the central region but 
very close to the two edges of the transparency window, and 
move away from the center when the transparency window 
becomes wider, as indicated by the arrows in Figs. \ref{fig4}(c) 
and (d) --- Wang et al. have already demonstrated this 
phenomenon in their recent experiments \cite{HaiWang}. 
Because the magnitude of ${\rm Im}\chi^{(1)}$ at the 
edges is much larger than at the center 
[see Fig. \ref{fig4} (e)], the magnitudes of the largest 
peaks of $\lambda$ of the three-level system are much smaller 
than those of the central peaks of $\lambda$ of the 
four-level system as seen in Figs. \ref{fig4}(a) and (b). In this 
calculation, we assume the dephasing rate $\gamma_{21}=0.01\gamma$. 
In this case the largest peak for the three-level system 
is more than one order smaller than the largest peak for the 
four-level system.

From the above comparison in terms of $\lambda$, we further confirm that the third-order susceptibility $\chi^{(3)}$ is indeed more effectively enhanced in the four-level system, and therefore the four-level system has an advantage for the realization of many quantum nonlinear optics phenomena. However, this conclusion does not mean the contribution of the three-level subsystem to the effectively enhanced $\lambda$ is negligible. In the next section, we will show the important influence of the quantum interference amongst the three contributions of light shifts to $\chi^{(3)}$ on the largest peak of $\lambda$ at a finite probe detuning when the transparency window becomes wider.

\section{Behavior of the largest peak of $\lambda$ of the 
four-level system} 
\begin{figure}[h]
\includegraphics[width=6cm, height=8cm, angle=270]{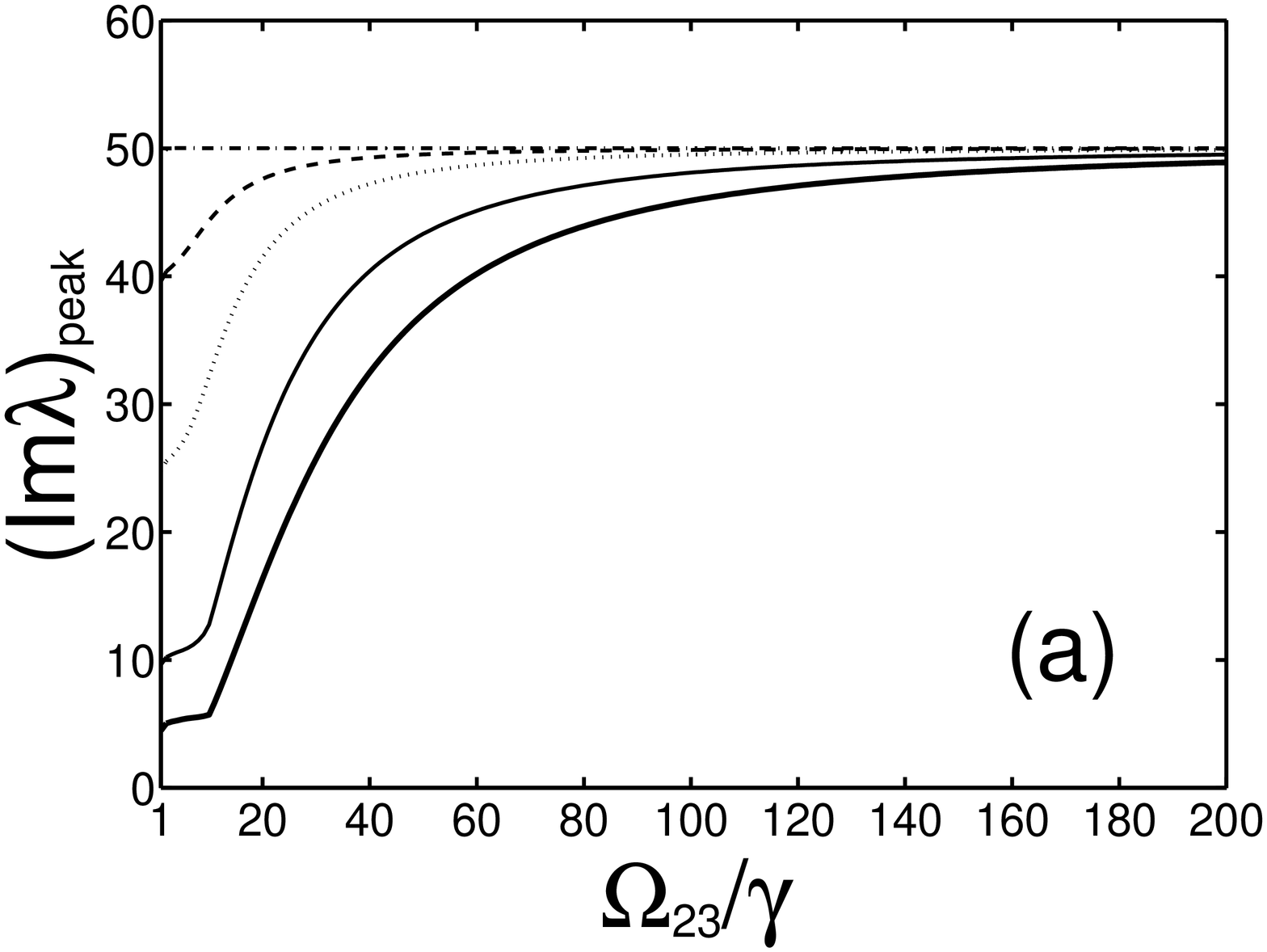}
\includegraphics[width=6cm, height=8cm, angle=270]{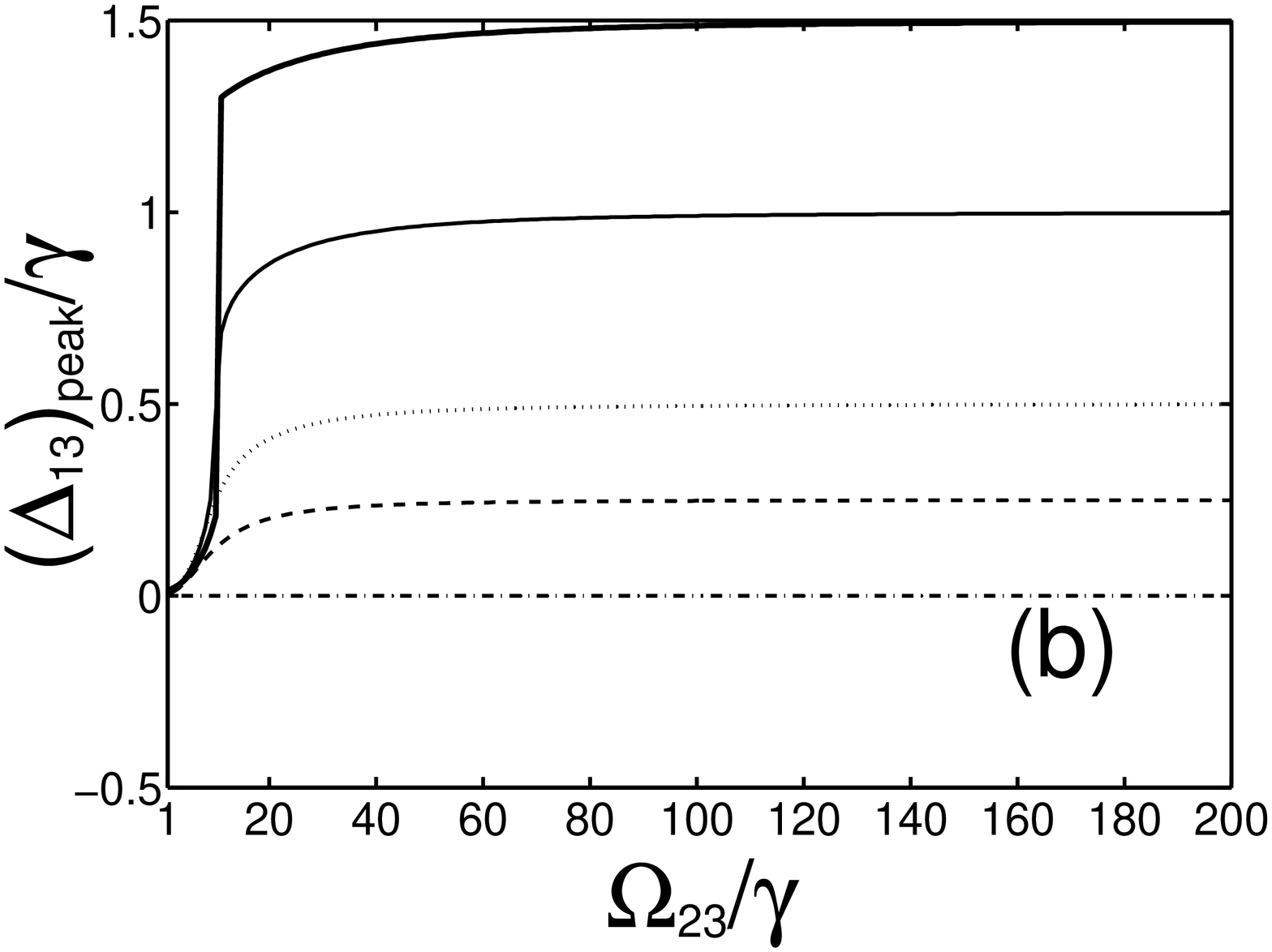}
\includegraphics[width=6cm, height=8cm, angle=270]{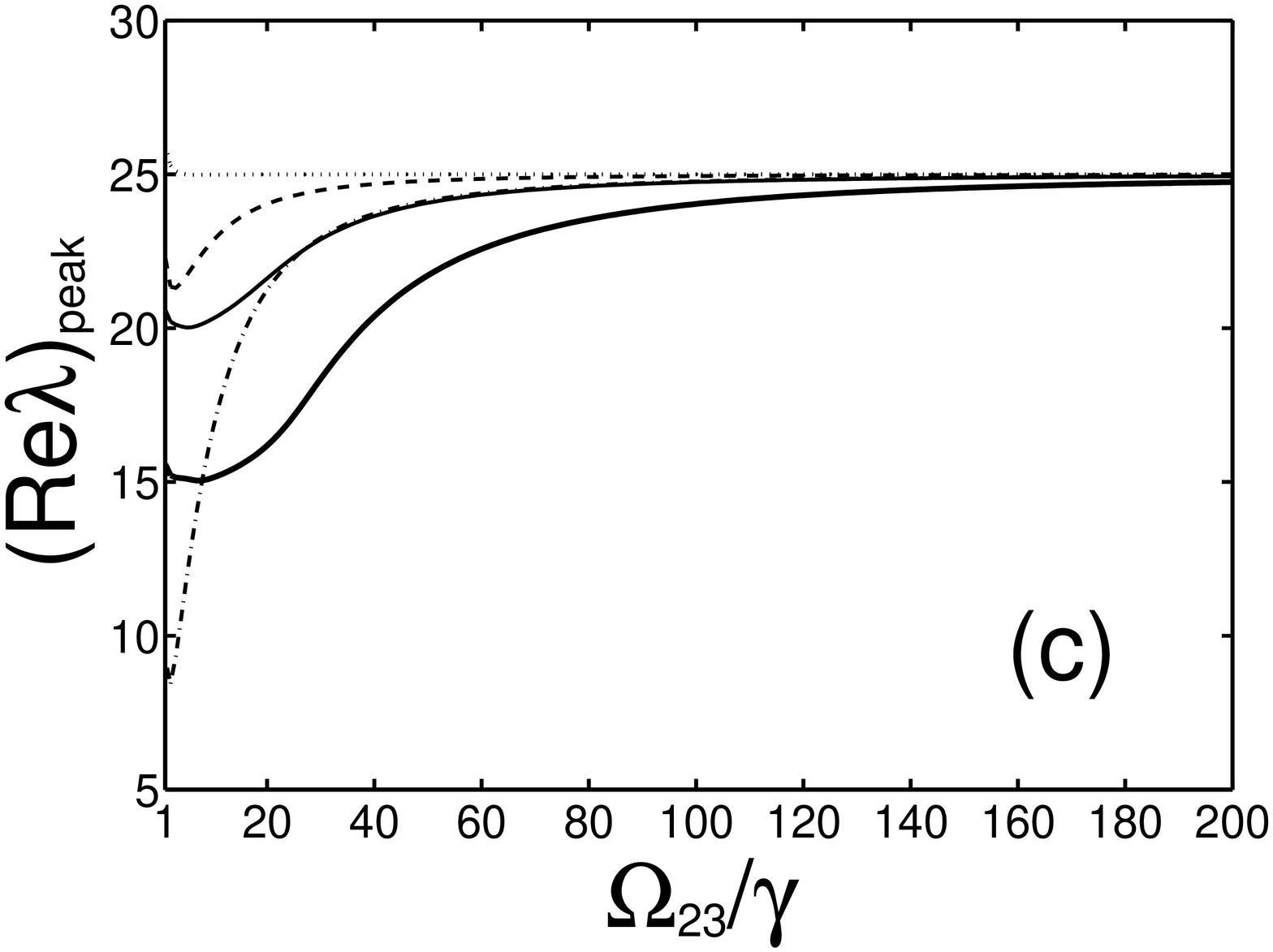}
\includegraphics[width=6cm, height=8cm, angle=270]{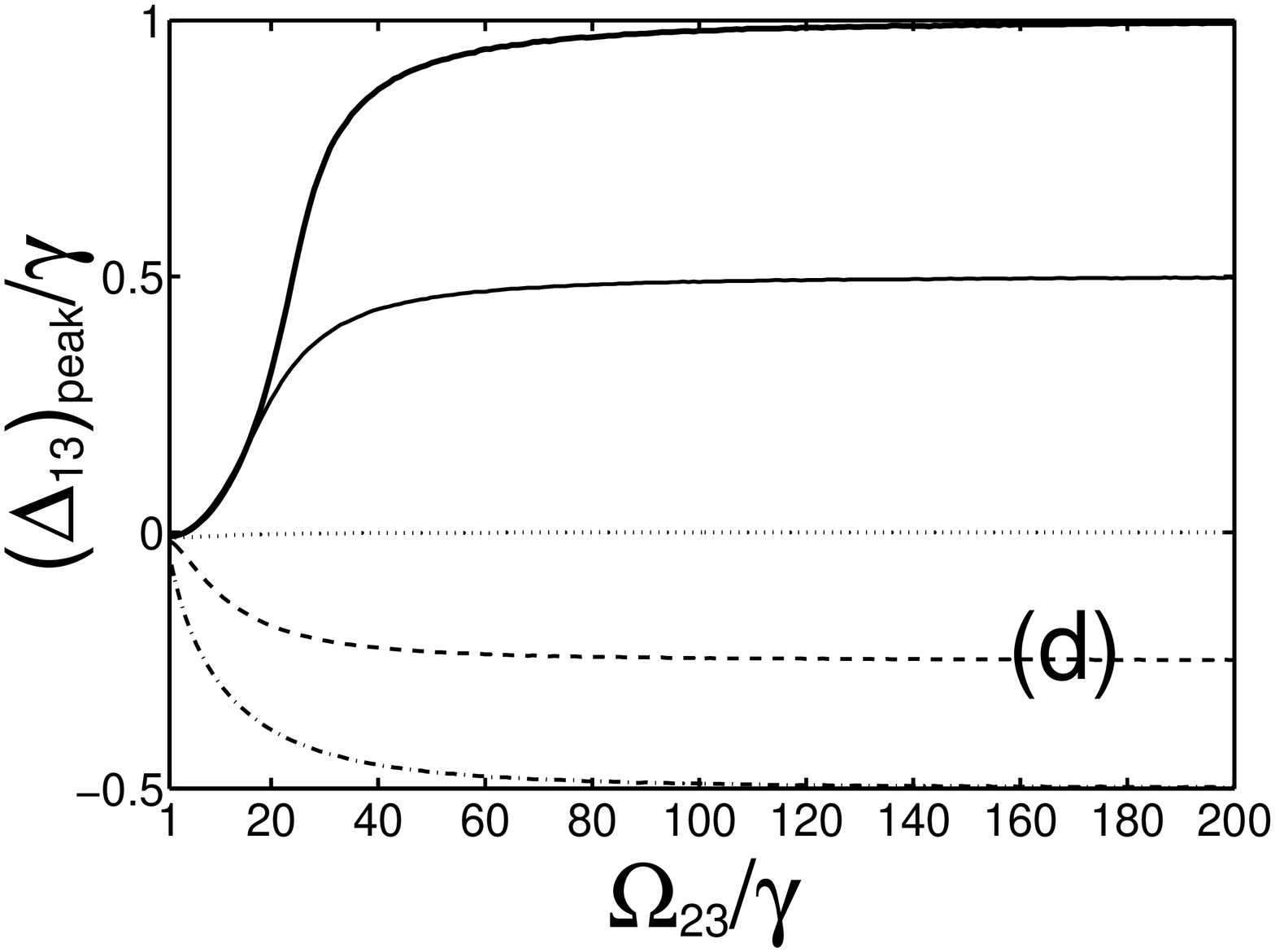}
\caption{ Value and detuning of the largest peak of $\lambda$ 
vs the Rabi frequency $\Omega_{23}$ and the atomic transition 
frequency difference $\omega_{31}-\omega_{42}$. 
(a) and (b) are for the imaginary part of $\lambda$. 
(c) and (d) are for the real part of $\lambda$.
The curves for $(\omega_{31}-\omega_{42})/\gamma=0$,
$-0.5$,$-1$,$-2$,$-3$ are 
plotted by dot-dashed lines, dashed lines, dotted lines, 
thin solid lines and thick solid lines, respectively. 
The ratio $\lambda$ is in units of $\mu^{2}/(\hbar \gamma)^{2}$. 
In this calculation,
 the dephasing rate $\gamma_{21}=0.01\gamma$.}
\label{fig5}
\end{figure}

In this section, we consider the detailed 
behavior of the largest peak of $\lambda$ of the four-level 
system with a small atomic transition frequency difference 
$\omega_{31}-\omega_{42}$ and a finite value of the dephasing 
rate $\gamma_{21}$. Because the dephasing rate $\gamma_{21}$ 
is always finite in reality \cite{Slowlight,EnhanceExperiment},
the absorption described by ${\rm Im}\chi^{(1)}$ is not zero 
when $\Delta_{13}=0\gamma$, and $\lambda$ is not divergent at 
this detuning. Thus, it is important in an implementation of 
the four-level scheme to determine under what condition the most 
effectively enhanced nonlinear susceptibility occurs under 
a finite dephasing rate $\gamma_{21}$. In Fig. \ref{fig4} (b), 
we can easily discern that the value of the largest peak 
(along the detuning axis) of ${\rm Im}\lambda$ increases when 
the coupling Rabi frequency $\Omega_{23}$ increases. By 
checking more carefully, we find that not only the value of 
the peak of $\lambda$ but also the position of the peak 
depends on the coupling Rabi frequency. This is depicted 
in Fig. \ref{fig5}. 

Fig. \ref{fig5}(a) shows the dependence of the value of 
the largest peak (along the detuning axis) of ${\rm Im}\lambda$, 
i.e., $({\rm Im}\lambda)_{peak}$, on the Rabi frequency, 
$\Omega_{23}$, and the atomic transition frequency 
difference $\omega_{31}-\omega_{42}$. In this calculation, 
we assume the dephasing rate $\gamma_{21}=0.01\gamma$. 
We find that as the coupling Rabi frequency, $\Omega_{23}$, 
increases, $({\rm Im}\lambda)_{peak}$ (for a 
fixed value of $\omega_{31}-\omega_{42}$) has a relatively 
small value at first, increases very rapidly, and then 
saturates to a relatively large value. For a fixed small 
value of $\Omega_{23}/\gamma$, $({\rm Im}\lambda)_{peak}$
 decreases monotonically as 
$(\omega_{31}-\omega_{42})/\gamma$ decreases from 
$0$ to $-3$. However, the saturation values for different 
values of $\omega_{31}-\omega_{42}$ always appear to 
be the same, and they are always larger than 
$({\rm Im}\lambda)_{peak}$ at small $\Omega_{23}/\gamma$.  
Although the value of the largest peak of ${\rm Re}\lambda$,
$({\rm Re}\lambda)_{peak}$, is not a 
monotonic function of $(\omega_{31}-\omega_{42})/\gamma$ 
at small $\Omega_{23}/\gamma$, its behavior is quite similar 
to that of $({\rm Im}\lambda)_{peak}$: as the coupling Rabi frequency 
$\Omega_{23}$ increases, $({\rm Re}\lambda)_{peak}$ 
always saturates at the same value, which is larger than its 
value at small $\Omega_{23}/\gamma$ in most cases, as shown 
in Fig. \ref{fig5}(c). The behavior of the largest peaks of 
${\rm Re}\lambda$ and ${\rm Im}\lambda$ indicate that if the 
most effectively enhanced nonlinear susceptibility is desired 
under a finite $\gamma_{21}$, then the coupling Rabi frequency 
$\Omega_{23}$ should be set as large as possible. This is 
especially true when the magnitude of the atomic transition 
frequency difference is not small.  

Next, from Fig. \ref{fig5}(b) [or (d)], we find that for a fixed value of $(\omega_{31}-\omega_{42})/\gamma$, when $({\rm Im}\lambda)_{peak}$ (or $({\rm Re}\lambda)_{peak}$) increases and saturates, the detuning $(\Delta_{13})_{peak}$, at which the largest peak occurs, also shifts asymptotically from $(\Delta_{13})_{peak}=0$ to another finite value.  The asymptotic value of the detuning is always equal to $-(\omega_{31}-\omega_{42})/2$, as shown in Fig. \ref{fig5}(b). This value exactly corresponds to the frequency of the probe light being resonant with the two-photon transitions between states $|1\rangle\leftrightarrow|4\rangle$, i.e.,  $2\omega_{p}=\omega_{31}+\omega_{42}$.  The asymptotic value of the detuning at which the largest peak of 
${\rm Re}\lambda$ occurs is always a half line width, $\gamma/2$, away from that of the largest peak of  ${\rm Im}\lambda$ for the same $(\omega_{31}-\omega_{42})/\gamma$, as shown in Fig. \ref{fig5}(d). 
This behavior of the largest peaks of ${\rm Re}\lambda$ and ${\rm Im}\lambda$ indicates that while the transparency window becomes wider, the constructive quantum interference amongst the three contributions of light shifts becomes more important, and the peak of $\lambda$ therefore shifts to a nonzero probe detuning. Thus if a most effectively enhanced nonlinear susceptibility is desired under a finite $\gamma_{21}$, the detuning of the probe light should be set close to the asymptotic values found in the above analysis. This is especially true when the magnitude of the atomic transition frequency difference is not very small.  Even if a certain ratio between ${\rm Re}\chi^{(3)}$  and ${\rm Im}\chi^{(3)}$ is required, the asymptotic values shown above will be important for determining the optimum detuning setting. Setting the detuning to zero, i.e., the center of the transparency window, might not be the best choice.

 Additionally, an interesting case occurs when 
$\omega_{31}-\omega_{42}=-\gamma$ 
(or $0\gamma$). In this case, $({\rm Re}\lambda)_{peak}$ 
(or $({\rm Im}\lambda)_{peak}$) reaches its 
saturation value for very small values of $\Omega_{23}$.  
This means that the enhancement of the nonlinear 
susceptibility can occur at very low coupling light intensity.

\begin{figure}[h]
\includegraphics[width=8cm, height=12cm, angle=270]{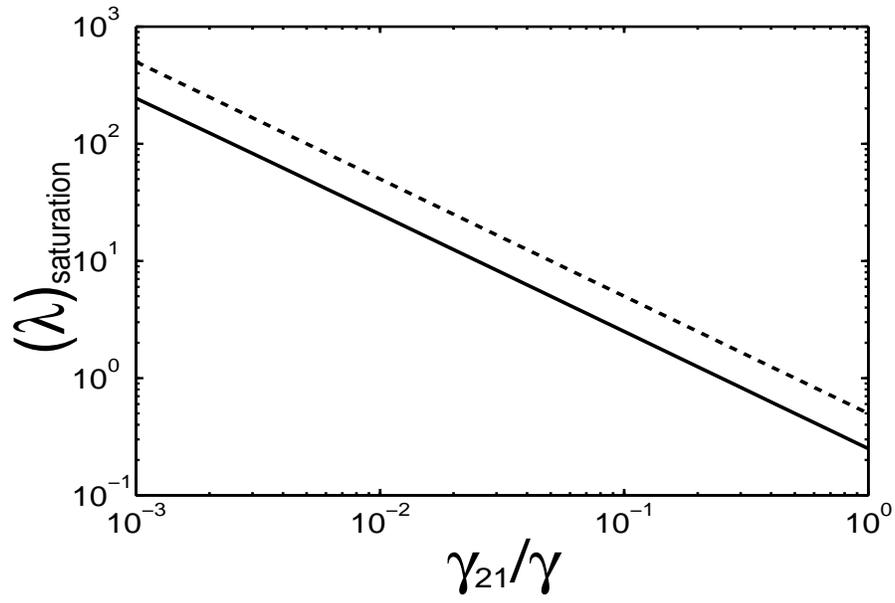}
\caption{Saturation value of the largest peak of $\lambda$ 
vs the dephasing rate $\gamma_{21}$. The ratio $\lambda$ is 
in units of $\mu^{2}/(\hbar \gamma)^{2}$. The solid line 
is for the real part of $\lambda$, 
and the dashed line for the imaginary part of $\lambda$.}
\label{fig6}
\end{figure}

In the calculations in Fig. \ref{fig4} and 
Fig. \ref{fig5},  we assumed the dephasing rate of the ground 
states is $\gamma _{21}=0.01\gamma $. More generally, 
we can calculate the dependence of the saturation values of the 
largest peaks of $\lambda$ on the dephasing rate $\gamma_{21}$, 
as shown in Fig. \ref{fig6}. If the dephasing rate $\gamma_{21}$ 
is much smaller than the decay rate $\gamma$ of the upper levels, 
then the four-level system can enhance the nonlinear susceptibility 
more effectively, as indicated by $\lambda$. This means that in 
order to employ the four-level system to more effectively enhance
the nonlinearity, one needs to reduce the dephasing rate of the 
ground states as much as possible. 

The analysis of the behavior of the largest peak of $\lambda$ also demonstrates the importance of the quantum interference amongst the three contributions of light shifts to $\chi^{(3)}$ when the transparency window is very wide. To implement the four-level system with a finite $\gamma_{21}$ to 
more effectively enhance the nonlinear susceptibility, the coupling 
Rabi frequency $\Omega_{23}$ should be set as large as possible and 
simultaneously the detuning of the probe light should be set as 
close as possible to the asymptotic values found in the above analysis. 
This is particularly important when the magnitude of the atomic 
transition frequency difference is not so small. This conclusion 
is very different from the proposal in Ref. \cite{ImamogluX3} 
due to the fact that in that work only the very special case when 
the dephasing rate vanishes, $\gamma_{21}=0$, was considered.    

\section{Conclusion}
We have studied in detail the third-order susceptibility for self-action of a four-level system, including the dephasing between the two ground states, under the condition of EIT by numerically solving the steady-state equations for the atomic density matrix. Through comparison of the four-level system with a three-level system with the same characteristic parameters, we discerned three major contributions from light shifts to the third-order susceptibility in the four-level system.  In particular, we found that quantum interference amongst the three contributions, which does not exist in the four-level system
with frequency-non-degenerate probes \cite{EITKerrXPM}, can not only enhance the third-order susceptibility more effectively in the four-level system than in the three-level system, but also make the ratio between its real part and imaginary part controllable. This unique feature means the four-level system has certain advantages for the realization of many quantum nonlinear optics phenomena. In implementing this scheme, it is important to note that in general the most effective enhancement of the nonlinear susceptibility does not occur exactly at the center of the transparency window. Instead, due to the constructive quantum interference, and a finite dephasing rate between the two ground states, the most effective enhancement occurs at an offset that is determined by the atomic transition frequency difference and the coupling Rabi frequency.

\newpage
\section{figure captions}

\newpage 

\newpage 

\newpage

\newpage 

\newpage 

\end{document}